# Near-Field Radiative Heat Transfer between Drift-biased Graphene through Nonreciprocal Surface Plasmons

Yong Zhang[1,2,*], Cheng-Long Zhou[1,2,*], Lei Qu[1,2], Hong-Liang Yi[1,2,†]

[1]*School of Energy Science and Engineering*, *Harbin Institute of Technology*, *Harbin 150001*, *P. R. China*
[2]*Key Laboratory of Aerospace Thermophysics*, *Ministry of Industry and Information Technology*, *Harbin 150001*, *P. R. China*

In this Rapid Communication, we theoretically demonstrate that near-field radiative heat transfer (NFRHT) can be modulated and enhanced by a new energy transmission mode of evanescent wave, i.e. the nonreciprocal surface plasmons polaritons (NSPPs). In addition to the well-known coupled surface plasmon polaritons (SPPs), applying a drift current on a graphene sheet leads to an extremely asymmetric photonic transmission model, which has never been noted in the noncontact heat exchanges at nanoscale before. The coupling of plasmons in the infrared bands dominates the NFRHT, associated with low loss (high loss and ultrahigh confinement) traveling along (against) the current. The dependence of NSPPs on the drift-current velocity as well as the vacuum gap is analyzed. It is found that the coupling of NSPPs at smaller and larger gap sizes exhibits different nonreciprocities. Finally, we also demonstrate that the prominent influence of the drift current on the radiative heat flux is found at a low chemical potential. These findings will open a new way to spectrally control NFRHT, which holds great potential for improving the performance of energy systems like near-field thermophotovoltaics and thermal modulator.

Due to the pioneering work of Polder and van Hove [1], it is well known that when two objects are close to each other (i.e., in the near-field gap), the radiative heat transfer (RHT) between them can be greatly enhanced [2-8]. The huge near field radiative heat transfer (NFRHT) allows many applications for energy conversion [9,10] and data carrier storage [11] as well as active noncontact thermal management at micro-nanoscale, such as thermal diodes [12-14], transistors [15-17], memories [18]. The coupling of evanescent modes plays a decisive role in huge heat transfer enhancement [19,20]. In particular, the NFRHT can be far ahead of the blackbody limit, either theoretically or experimentally, via the resonant coupling of surface phonon polaritons (SPhPs) [21] or surface plasmon polaritons (SPPs) [22,23]. Recently, excitations of hyperbolic polaritons [24,25], magnetic polaritons [3], ellipse polaritons [26] and epsilon-near-zero modes [27] with various model of metamaterials have also been reported to further enhance or tune the NFRHT. Moreover, compared to the case of passive field, the near-field radiative transport in the presence of a static magnetic field can be further modulated due to magneto-optical effect [28] and magneto-plasmon polaritons (MPP) modes [29,30]. These strategies have largely developed methods for regulating and enhancing near-field thermal radiation through the interaction of surface polaritons.

Here, based on the unusual nonreciprocal and diffractionless properties of surface plasmon polaritons propagating, we propose a new coupling of evanescent modes for NFRHT, i.e., the nonreciprocal surface plasmon polaritons (NSPPs). The NSPPs have broken the Lorentz reciprocity principle, making it possible to dynamically control and collimate the direction of these waves supported by manipulating the nonreciprocity strength and modifying the available states [31]. Even though it is difficult to obtain a sufficiently large drift velocity required for a strong nonreciprocal response in most semiconductors and metals [32,33], graphene obtain excellent NSPPs by applying drift-current bias to the host surface thanks to its ultrahigh electron mobility (with a high Fermi velocity $v_f \approx 10^8$cm/s) [34]. Graphene, a novel natural two-dimensional layered material, has been attracting significant research attention due to its outstanding electronic and optical properties and tunable bandgaps. Ilic first proposed the graphene-graphene near-field heat transfer controlled by plasma in the infrared band and the model of graphene radiative heat to electricity energy conversion [35,36]. In this Rapid Communication, we will theoretically investigate the possible effect of NSPPs on the NFRHT between two monolayer graphene sheets with a drift current.

---

[*] These authors contributed equally to this work.
[†] Corresponding author. E-mail address: yihongliang@hit.edu.cn

As depicted in Fig. 1, two monolayer graphene sheets are brought into close proximity with a vacuum gap of $d$, where a longitudinal voltage $V_{DC}$ induces a drifting of electrons along the sheet with velocity $\hat{v}_d = v_d \hat{y}$. For simplicity, we firstly assume that the two bodies are mirror images of each other, that is, manipulated by the same direction of the drift-current bias. Considering the self-consistent quantum mechanical methods and ignoring the direct effect of the drift current velocity on the energy of charges, Refs. [37] and [38] recently demonstrated that the graphene's conductivity with this drift becomes nonlocal, and can be written as

$$\sigma_d(v_d, k_y) = \frac{\omega}{\omega - k_y v_d} \sigma_g(\omega - k_y v_d) \quad (1)$$

where $\omega$ and $k_y$ denote the angular frequency and the wavevector component along the drift-direction, respectively. Intuitively, the factor of $k_y v_d$ is a Doppler shift introduced by the drift bias [37]. And $\sigma_g$ is a rigorous conductivity model [39] that incorporates the intrinsic nonlocal response of graphene into the frequency band using the Bhatnagar-Gross-Krook (BGK) approach. To date, in the past studies of such plasmons, it has been assumed that it is invariable in the transverse plane axis, thus considering drift-biased graphene as a 2D waveguide problem [40].

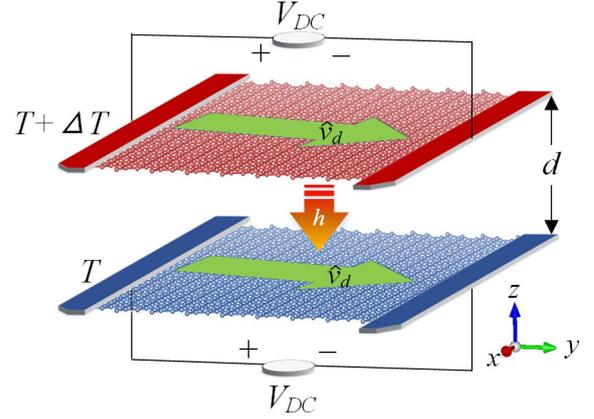

FIG. 1 Schematic of NFRHT between two monolayer graphene sheets with a drift current $\hat{v}_d = v_d \hat{y}$ in the y-direction of the plane.

We maintain the two bodies at the temperatures of $T + \Delta T$ and $T$, respectively. Since graphene is a good conductor, it will not affect the temperature distribution of the system due to Joule heating effect by the drift current. In the framework of fluctuation electrodynamics, the heat transfer coefficient (HTC) between two graphene sheets is given by [1]

$$h(T,d) = \int_0^\infty h_\omega(\omega) d\omega = \int_0^\infty \frac{\partial \Theta(\omega,T)}{\partial T} \Phi(\omega) d\omega \int_{-\infty}^\infty \int_{-\infty}^\infty \frac{\xi(\omega,k_x,k_y)}{8\pi^3} dk_x dk_y \quad (2)$$

where $\Theta(\omega,T) = \hbar\omega/[\exp(\hbar\omega/k_B T) - 1]$ is the mean energy of a Planck oscillator at angular frequency $\omega$ and temperature $T$, in which $T$ is the absolute temperature that we assume equal to 300 K (room temperature) throughout this work. $\Phi(\omega)$ is the spectral energy transfer function given by

$\Phi(\omega) = \int_{-\infty}^\infty \int_{-\infty}^\infty \frac{\xi(\omega,k_x,k_y)}{8\pi^3} dk_x dk_y$. $\xi(\omega,k_x,k_y)$ is the photonic transmission coefficient (PTC) that describes the probability of two thermally excited photons, which can be written as [28, 29]

$$\xi(\omega,k_x,k_y) = \begin{cases} \mathrm{Tr}[(\mathbf{I} - \mathbf{R}_2^* \mathbf{R}_2) \mathbf{D} (\mathbf{I} - \mathbf{R}_1 \mathbf{R}_1^*) \mathbf{D}^*], & \beta < \beta_0 \\ \mathrm{Tr}[(\mathbf{R}_2^* - \mathbf{R}_2) \mathbf{D} (\mathbf{R}_1 - \mathbf{R}_1^*) \mathbf{D}^*] e^{-2|k_z|d}, & \beta > \beta_0 \end{cases} \quad (3)$$

for propagating ($\beta < \beta_0$) and evanescent ($\beta > \beta_0$) waves where $\beta=\sqrt{k_x^2 + k_y^2}$ is the surface parallel wavevector and $\beta_0=\omega/c$ is the wavevector in vacuum. $k_z=\sqrt{\beta_0^2 - \beta^2}$ is the tangential wavevector along $z$ direction in vacuum and * signifies the complex conjugate. The 2×2 matrix **D** is defined as $\mathbf{D}=(\mathbf{I}-\mathbf{R}_1\mathbf{R}_2 e^{2ik_z d})^{-1}$ which describes the usual Fabry-Perot-like denominator resulting from the multiple scattering between the two interfaces. The reflection matrix **R** is a 2×2 matrix in the polarization representation (for the numerical method of reflection matrix **R**, see SI).

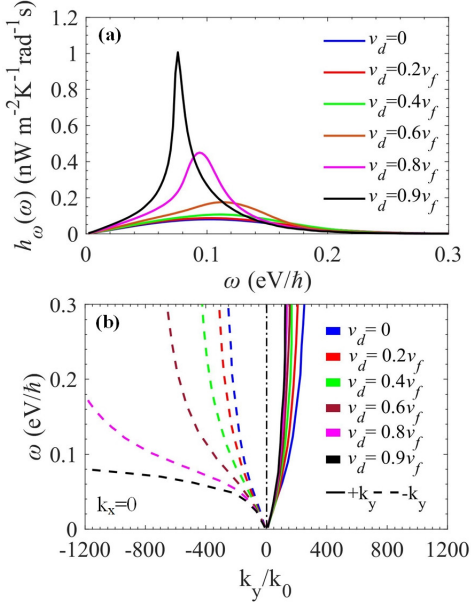

FIG. 2. (a) Spectral HTC as a function of the frequency for a vacuum gap $d = 10$ nm and a low chemical potential $\mu = 0$ eV. The different lines correspond to different drift current velocities. (b) Dispersion of SPPs with respect to the frequency for $-k_y$ (dashed lines) and $k_y$ (solid lines) in the drift-biased graphene sheet.

To visualize the contribution of the nano-structure to the near-field radiative heat transfer, Figure 2(a) presents the spectral HTC between two graphene sheets with a lower chemical potential of $\mu = 0$ eV at vacuum gap distance of $d = 10$ nm with different drift-current velocity $v_d$ from 0 to 0.9 $v_f$. This spectral HTC is defined as the HTC per unit of frequency or photon energy [4]. Notice that the maximum of spectral HTC increases drastically with the drift-current velocity, reaching a maximum at $v_d$ = 0.9 $v_f$. When the drift velocity is less than 0.4 $v_f$, the frequency of peak remains basically unchanged. However, when the velocity exceeds 0.6 $v_f$, the maxima of the HTC is redshifted from 0.114 eV/$\hbar$ for $v_d = 0.6$ $v_f$ to around 0.076 eV/$\hbar$ for $v_d = 0.9$ $v_f$. These results illustrate the high tunability of NFRHT via the drift-current velocity.

Figure 2(b) shows the dispersion relations of SPPs for $-k_y$ (dashed lines) and $+k_y$ (solid lines) in the drift-biased graphene sheet. We see that for a larger $v_d$, the $-k_y$ branch is dragged to a higher wave vector and meanwhile compressed to a lower frequency. Eventually the $-k_y$ branch for $0.9v_f$ is parallel to the $k_y$ axis, resulting in the lack of nonreciprocal SPPs along $-k_y$ at a large frequency. The above analysis is consistent with our observation about the redshift and enhancement of the spectral HTC as depicted in Fig. 2(a).

To confirm that NSPPs are indeed responsible for the amplification of NFRHT in our structure, we have analyzed the photonic transmission coefficient. With different drift-current velocity (0, 0.6$v_f$, and 0.9$v_f$), the contour plots of PTC at frequency of the spectral HTC's peak are shown in Fig. 3(a)-(c). It is well known that the graphene sheet supports isotropic SPPs due to collective charge oscillations coupled to light [41]. In Fig. 3(a), the bright band and its corresponding dispersion relationship (green dotted line) well reflect the isotropic SPP of ordinary graphene. However, in the presence of voltage bias, the collective charge are strongly affected by the dragging effect of these drifting charges, which causes guided waves to effectively show asymmetric effect, in turn presents the NSPPs [40]. As one can see from Fig. 3(b) and (c), the symmetry of eigenstates with positive and negative $k_y$ is damaged by the drift bias, leading to a strongly anisotropic two-dimensional surface in which the Poynting vectors and wavevector are no longer aligned uniformly [40]. Meanwhile, these drift charges with higher kinetic energies would easily drag the SPPs generated by the collective charge to a larger wave vector range, thereby exciting a higher maxima in the spectral HTC in Fig. 2(a). When the drift current reaches 0.9 $v_f$, as shown in Fig. 3(c), the maximum wavevectors of the bright branches in the bottom $k_y$ quadrants can approach -1200 $k_0$, while the maximum wavevectors of positive $k_y$ quadrants just stay at around 160 $k_0$. The bright bands in the negative and positive quadrants correspond to ultraconfined and lossy NSPPs (opposite the drift current) and low-loss plasmons (along the drift current), respectively [40].

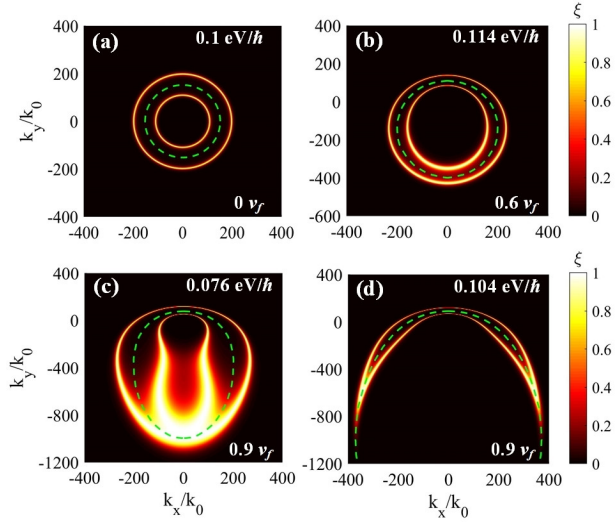

FIG. 3. Photonic transmission coefficient at drift current velocity of (a) 0, (b) 0.6 $v_f$, and (c) 0.9 $v_f$ with frequency of 0.07 eV/$\hbar$. (d) Photonic transmission coefficient at frequency of 0.104 eV/$\hbar$ with a drift current velocity of 0.9 $v_f$.

These drifting charges also result in a negative Landau damping, in which SPPs gains kinetic energy from the drifting charges and thus are magnified. Thus, plasmons in the bottom $k$ quadrant (propagating towards these drifting charges) are significantly more unconstrained and brighter than their positive counterparts in Fig. 3(c). Meanwhile, a larger kinetic energy of charges (lager drift velocity of charges) would produce a brighter bright band in the bottom $k$ quadrant in Figs. 3(b) and 3(c), further leading to a positive correlation between the maximum value of the spectrum HTC and the drift velocity of Fig. 2(a). To confirm that NSPPs motivated by these drifting charges are indeed responsible for the NFRHT, we also show the dispersion relations in Fig. 3(b) and (c), denoted by the dotted line, which is obtained by calculating the dispersion formulation for anisotropic materials (Eq. S17 in SI). This green dotted line nicely locates between the two bright branches, which unambiguously demonstrates that NSPPs dominate the NFRHT in our system. Due to the asymmetric feature of the NSPPs in the wavevector space, one can expect to tune the NFRHT by change the relate orientation between the two graphene sheets (See in SI).

In addition, as the drift current velocity gradually increases, spectral HTC curve appears a sharper peak in Fig. 2. For $v_d = 0.9\ v_f$, we can see the spectral HTC quickly reach as high as 1.05 nW·m$^{-2}$·K$^{-1}$·rad$^{-1}$·s at 0.076 eV/$\hbar$, and decreases rapidly as frequency $\omega$ further increases. In order to expound this result, we plot the PTC at a frequency of 0.104 eV/$\hbar$ in Fig. 3(d). We can clearly see the lack of unidirectional wavevector appears in the bottom $k_y$ quadrants, i.e., NSPPs are no longer closed. The open shape of NSPPs appears due to the intrinsic nonlocal response of graphene, as the finite velocity of electrons $v_f$ cannot follow the increasingly quick variations of the plasmons, a behavior consistent with the case of nonreciprocal plasmons on metal-dielectric interfaces biased with a magnetic field [42]. In Fig. 3(d), the open dispersion curve (green dotted line) also corresponds to the lack of NSPPs at large $k_y$ of bottom quadrants.

In Fig. 4(a) we plot the HTC for the NSPPs system as considered in Fig. 1, where we vary the vacuum gap while keeping all other parameters fixed. We also show the progressive enhancement of HTC stimulated by increased drift-current velocity in fixed vacuum gap, and the enhancement factor of HTC is shown in the inset of Fig. 4(a). For a high drift current parameters, the system shows a significant enhancement, especially in the regime of low vacuum gap, where the enhancement can be up to more than 4-fold. We also show PTC along $y$ axis [$\xi(\omega,\ 0,\ k_y)$] in Fig. 4(b) for different vacuum gaps.

At a small gap size, there are bright bands at $-k_y$ region with a large value of wavevector, thereby enhancing the HTC as shown in Fig. 4(a). This is because that the two graphene sheets are so close to each other that high $k_y$ evanescent waves barely decay before reaching the surface, and once reaching the surface, they are coupled to each other to produce large heat fluxes.

While for a large $d$, the bright bands in $-k_y$ disappear, which means that the NSPPs against the direction of drift current is filtered by the free space due to their ultra-confined and lossy character. Only the NSPPs towards $+k_y$ contribute to the heat transfer. This is also consistent with the results in Fig. 3. Fig. 4(c) gives the PTC at $d = 100$ nm. We can observe that the bright band is concentrated in the $+k_y$ quadrant which is different from those in Figs. 3(b)-(d). Due to the small wavevector of the $+k_y$ branch at all the drift velocities, the $v_d$ makes negligible impact on the NFRHT for the two graphene sheet separated at a large gap size as shown in Fig. 4(a).

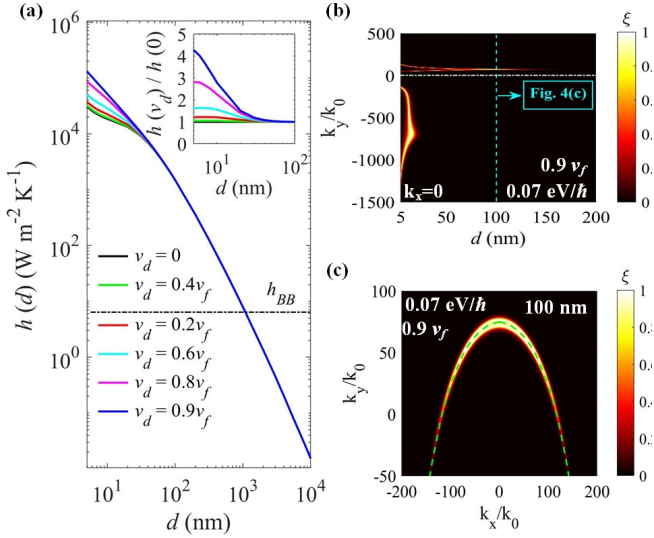

FIG. 4. (a) HTC as a function of the vacuum gap. The different lines correspond to different drift current velocity. The enhancement factor between the bias system and the zero-bias system is shown in the inset. The black dash-dotted line is the RHTC of the black body given by $h_{BB} = 4\sigma_{SB}T^3 \approx 6.1$ Wm$^{-2}$K$^{-1}$, where $\sigma_{SB}$ is the Stefan-Boltzmann constant. (b) Photonic transmission coefficient along $y$ axis for different vacuum gaps. (c) Photonic transmission coefficient at vacuum gap of 100 nm. $\omega = 0.07$ eV/$\hbar$ and $v_d = 0.9\ v_f$ are considered in (b) and (c).

Meanwhile, the presence of chemical potential $\mu$ also play a nonnegligible role in NFHTC of graphene's NSPPs. Such a dependence of room-temperature HTC controlled by the different chemical potentials on drift current velocity is shown in Fig. 5. As seen in Fig. 5(a), we observe the positive correlation between HTC and drift current velocity is greatly suppressed with the increasing of chemical potential, especially for $\mu \geq 0.2$ eV. The results demonstrate that the influence of NSPPs on the HTC is mainly concentrated on the low chemical potential. To observe this effect more intuitively, in the inset of Fig. 5(a), we plot the ratio of HTC between the zero-bias system and bias system. For the zero chemical potential system, the maximum ratio can be as high as 3-fold. In contrast, the ratio for the black line representing high chemical potential is always close to 1.

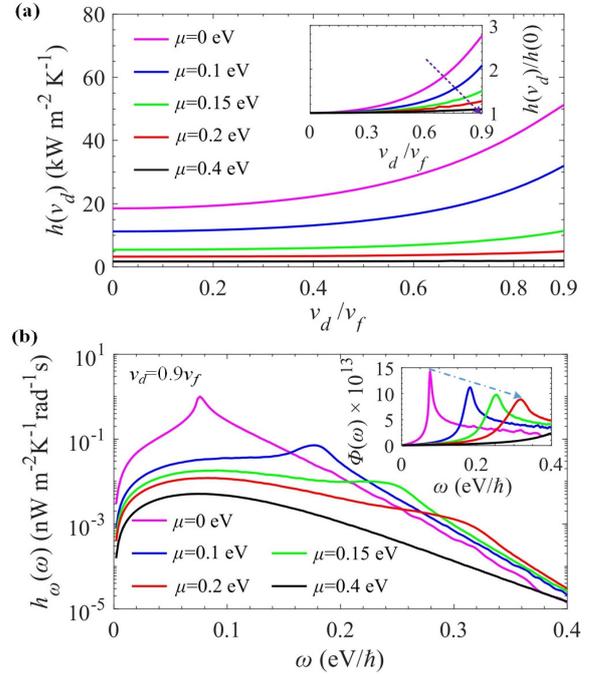

FIG. 5. (a) HTC as a function of drift current velocity. The different lines correspond to different chemical potentials. The inset shows the enhancement factor of HTC for the bias system with respect to the zero-bias system. (b) Spectral HTC as a function of the frequency. The inset shows the energy transfer function $\Phi(\omega)$ [43] as a function of the frequency. The different lines correspond to different chemical potentials.

In Fig. 5(b), we calculate the spectral energy transfer function $\Phi(\omega)$ and the spectral HTC $h_\omega(\omega)$, given by [43] for different chemical potentials with a high drift current velocity of 0.9 $v_f$. As the chemical formula increases, the NSPPs are excited only by high photonic energy, which corresponds to the blue-shifted peak of energy transfer function shown in the inset of Fig. 5(b). This is also reflected in the contour plots of PTC in SI. However, the contribution of NSPPs to HTC at high frequencies are negligible, decaying exponentially at room temperature. Therefore, in Fig. 5(b), when the chemical potential is 0.4eV, it is difficult to observe the peak of the spectral HTC caused by NSPPs.

In summary, we have proposed a novel energy transmission mechanism of evanescent wave to enhance and dynamically control NFRHT based on the strong unidirection and high tunability of NSPPs.

These NSPPs are excited by drift current on the graphene sheet. We theoretically prove that applying a drift current to a graphene sheet results in an extremely asymmetric modal dispersion and photonic transmission mode, which is associated with low-loss SPPs along the drift direction and lossy SPPs in the opposite direction. In addition, the NSPPs can exhibit an interesting dependency with the NFRHT on drift current velocity, nonlocal effects, vacuum gap and chemical potential. The fundamental understanding gained here will open a new way to spectrally tune near-field radiative heat transfer between metamaterials for energy conversion and thermal management.

This work is supported by the National Natural Science Foundation of China (Grant No. 51706053), as well as the Fundamental Research Funds for the Central Universities (Grant No. HIT. NSRIF. 201842), and by the China Postdoctoral Science Foundation (Grant No. 2017M610208). The authors particularly thank Dr. D. Y. Xu, Dr. T. F. Li and Dr. X. P. Luo at Harbin and Prof. J. S. Gomez-Diaz at California for helpful discussions.

C. L. Zhou provided simulations, analysis and the manuscript. Y. Zhang provided idea and feedback on the manuscript.